\documentclass[american,aps,longbibliograph,reprint]{revtex4-2}
\usepackage[T1]{fontenc}
\usepackage[utf8]{inputenc}
\usepackage{colortbl}
\usepackage{array}
\usepackage{multirow}
\usepackage{amsmath}
\usepackage{amssymb}
\usepackage{graphicx}

\makeatletter

\providecommand{\tabularnewline}{\\}
\usepackage[unicode=true,breaklinks=true]{hyperref}
\hypersetup{colorlinks}

\makeatother

\usepackage{babel}
\begin{document}
	\title{Few Dipolar Bosons in a Ring Trap: Quantum States and Transitions
		between Them}
	\author{Jieli Qin}
	\email{qinjieli@126.com; 104531@gzhu.edu.cn}
	
	\affiliation{School of Physics and Materials Science, Guangzhou University, Guangzhou
		510006, China}
	\begin{abstract}
		The degenerate dipolar Bose gas has attracted great research interest
		for studying novel quantum states of matter due to the interplay between
		short-range contact and long-range dipole-dipole interactions. The
		superfluid, quantum droplet, and supersolid states,
		along with the transition between them, have been widely studied using
		the extended mean-field theory, which relies on the Lee-Huang-Yang
		correction to approximately account for the effect of quantum fluctuation.
		In this work, we examine the system of a few dipolar bosons trapped
		in a ring potential employing the exact diagonalization method, which
		can yield the full quantum nature of the system. By examining the
		quantum correlation and coherence properties of the system, we reveal
		a rich variety of quantum states (including Wigner crystal, superfluid,
		quantum droplet, and the mixed states of them) 
		and their formation mechanisms. Furthermore, we show that
		transitions between these quantum states can be induced by tuning
		the interaction strength and particle number. The findings of this
		work can provide valuable insights into the dipolar gas at the microscopic
		quantum level, and may also find applications in the fields such as quantum
		sensing and quantum information.
	\end{abstract}
	\maketitle
	
	\section{Introduction}\label{sec:Introduction}
	
	The coexistence of short-range contact interaction (CI) and long-range
	dipole-dipole interaction (DDI) can significantly enrich the quantum
	physics of degenerate dipolar Bose gas, making it a fascinating platform
	for studying novel quantum states of matter \citep{lahaye_physics_2009,bottcher_new_2020,chomaz_dipolar_2022}.
	More than two decades ago, dipolar atomic Bose-Einstein condensate
	(BEC) has been realized in the laboratory \citep{griesmaier_bose-einstein_2005},
	and its unique superfluid (SF) properties are revealed later \citep{wilson_critical_2010,ticknor_anisotropic_2011,wenzel_anisotropic_2018}.
	In recent years, dipolar molecular BEC has also achieved rapid development
	\citep{langen_quantum_2024,zeng_towards_2025}. In addition to dipolar
	BECs, dipolar quantum droplet (QD) \citep{kadau_observing_2016,ferrier-barbut_observation_2016}
	and dipolar supersolid (SS) \citep{tanzi_observation_2019,bottcher_transient_2019,chomaz_long-lived_2019}
	have been experimentally observed as well, and are receiving extensive
	research attention nowadays. Moreover, the Wigner crystal (WC) state
	has also been theoretically predicted to exist in systems of dipolar
	bosons \citep{mitra_hexatic_2009,deuretzbacher_ground-state_2010,cremon_tunable_2010,xu_wigner_2012}.
	
	The dipolar Bose gases are usually theoretically modeled using an
	extended Gross-Pitaevskii equation (GPE) with a Lee-Huang-Yang (LHY)
	correction term included to approximately account for the effect of
	quantum fluctuation. There are a great many works employing this
	method. For a few examples of them, see Refs. \citep{wachtler_quantum_2016,wachtler_ground-state_2016,baillie_self-bound_2016,bisset_ground-state_2016,edler_quantum_2017,boudjemaaQuantumDiluteDroplets2017,blakie_supersolidity_2020,zhao_elastic_2025}.
	This method can well describe systems that contain a large number
	of particles, and explain phenomena such as QD and SS at the macroscopic
	quantum scale. However, in essence, it is an extended mean-field method,
	thus cannot capture the full quantum nature of the system. Therefore,
	it is totally unable to describe the WC states, and also unable to
	reveal the microscopic quantum origin of SF, QD and SS states, and
	their quantum depletion and correlation properties. To better capture
	the quantum nature of the system, methods such as multiconfigurational
	time-dependent Hartree \citep{chatterjee_ultracold_2013,chatterjee_order_2018,chatterjee_correlations_2019,chatterjee_detecting_2020,royPhasesCoherenceStrongly2024,molignini_interaction_2025,molignini_stability_2025,chakrabarti_out--equilibrium_2026},
	density matrix renormalization group \citep{biedron_extended_2018,kraus_superfluid_2020,kraus_quantum_2022,korbmacher_transversal_2023,tuzemen_diminished_2025},
	Monte-Carlo \citep{sanchez-baena_few_2018,dornheim_path-integral_2020,ghosh_path_2024,bombin_quantum-monte-carlo--based_2024,bombin_creating_2025}
	and exact diagonalization \citep{deuretzbacher_ground-state_2010,cremon_tunable_2010,xu_wigner_2012,oldziejewski_roton_2018,oldziejewski_strongly_2020,tuzemen_diminished_2025}
	could be applied. Among them, the exact diagonalization method yields
	the numerically exact eigenenergies and eigenstates of a quantum system
	without introducing any systematic approximations. Its precision is
	only limited by the truncation of the Hilbert space, at the cost of
	only being able to cope with systems containing very few particles.
	
	In this work, we study a system of a few dipolar bosons trapped within
	a ring potential. We obtain the ground state of the system using the
	exact diagonalization method. We show that under different particle
	numbers and interaction strengths, the system exhibits various types
	of quantum states, for instance, WC, SF, QD, WC-QD mixed and SF-QD
	mixed states. We further examine the transitions between these quantum
	states when tuning the interaction strength and particle number of
	the system. We find that quantum fluctuation plays an essential role
	when the particle number is small. As the particle number increases,
	the quantum results gradually approach the mean-field results. The
	findings of this work reveal new microscopic features of the dipolar
	bosons system, and deepen our understanding of the dipolar Bose gas
	at the full quantum level. Noting that systems of dipolar bosons have
	been proposed to build various types of atomtronic devices \citep{maik_dipolar_2011,wilsmann_control_2018,w_controlling_2023,rovirola_ultracold_2024,ymai_tunneling_2025,wittmann_w_atomtronic_2026},
	the results of this work might also find applications in quantum technologies.
	
	The rest of this paper is organized as follows. In Sec. \ref{sec:Model},
	we describe the physical model we are studying in this work, and present
	the Hamiltonian of the system. In Sec. \ref{sec:MeanFieldResults},
	we briefly examine the system using the mean-field method, and show
	the superfluid to bright soliton phase transition of dipolar gas.
	These mean-field results offer a benchmark for manifesting the quantum
	feature of the system. Sec. \ref{sec:QuantumResults} shows the main
	results of this paper. In this section, we study the system in the
	full quantum way, and we compare the quantum results with their mean-field
	correspondence to reveal the quantum nature of the system. At last,
	our main conclusions are drawn in Sec. \ref{sec:Conclusions}.
	
	\section{Model}\label{sec:Model}
	
	\begin{figure}
		\begin{centering}
			\includegraphics{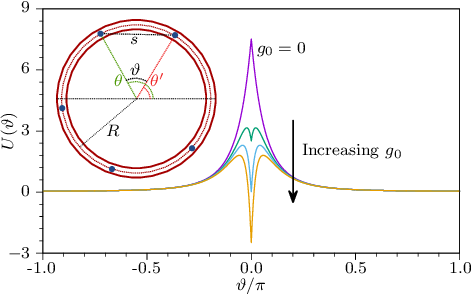}
			\par\end{centering}
		\caption{Diagram of the ring trapped few dipolar particles system (inset) and
			the interaction potential between two dipolar particles. The radius
			of the ring trap is $R$. The angular and linear distances between
			two particles located at angular positions $\theta$ and $\theta'$
			are labeled as $\vartheta$ and $s$. The interaction potential between
			these two particles is $U\left(\vartheta\right)$, given by Eq. (\ref{eq:Interaction}),
			and it is plotted for $g_{d}=3.0$ and varying $g_{0}$. The violet
			line corresponds to $g_{0}=0$, i.e., it is the plot of $g_d U_{d}\left(\vartheta\right)$.
			For the lines from top to bottom, $g_{0}$ increases. In plotting
			this figure, the Dirac delta function is approximated by $\delta\left(x\right)\approx\Delta\left(x\right)=5\left[x+\mathrm{sgn}\left(x\right)/2\right]^{-6}/64$.
			Since the long-range DDI potential $U_{d}\left(x\right)$ approximately
			decays in a manner of $1/x^{3}$, this $1/x^{6}$-decaying function
			can be regarded as short-range to approximate the delta function.
			We emphasize that in the numerical calculations, this approximation
			is not used; we deal with the delta function exactly.}\label{fig:DiagramAndInteractionU}
	\end{figure}
	
	We consider a system of identical dipolar Bose particles trapped in
	a ring potential with radius $R$ and transverse length scale $a_{\bot}$.
	For a diagram of the system, see the inset of Fig. \ref{fig:DiagramAndInteractionU}.
	We assume that the transverse trap is very tight, such that $a_{\bot}\ll2\pi R$
	is fulfilled. Under this assumption, the transverse excited modes
	are barely populated; therefore, the system can be reduced to one
	dimension, being described by the following Hamiltonian \citep{zollner_ground_2011,zollner_bosonic_2011}
	\begin{align}
		H &=-\frac{\hbar^{2}}{2mR^{2}}\int\hat{\psi}^{\dagger}(\theta)\frac{\partial^{2}}{\partial\theta^{2}}\hat{\psi}(\theta)d\theta \nonumber \\ &+\frac{1}{2}\int\hat{\psi}^{\dagger}(\theta)\hat{\psi}^{\dagger}(\theta^{'})U(\vartheta)\hat{\psi}(\theta^{'})\hat{\psi}(\theta)d\theta'd\theta,\label{eq:Hamiltonian}
	\end{align}
	where $\hat{\psi}\left(\theta\right)$ is the bosonic field operator
	with $\theta$ specifying the particle's angular position in the ring,
	$\hbar$ is the reduced Planck constant, $m$ is the mass of the particle,
	$U\left(\vartheta\right)$ with $\vartheta=\theta-\theta'$ is the
	interaction potential between two particles located at angular positions
	$\theta$ and $\theta'$ in the ring. Since we are considering dipolar
	particles, $U\left(\vartheta\right)$ is split into two parts ---
	a long-range DDI part and a short-range CI part,
	\begin{equation}
		U\left(\vartheta\right)=g_{d}U_{d}\left(\vartheta\right)-g_{0}\delta\left(\vartheta\right),\label{eq:Interaction}
	\end{equation}
	with
	\begin{equation}
		U_{d}\left(\vartheta\right)=\sqrt{2\pi}\left(1+s^{2}\right)e^{s^{2}/2}\mathrm{erfc}\left(\left|s\right|/\sqrt{2}\right)-2\left|s\right|.\label{eq:LongRangeInteraction}
	\end{equation}
	Here, the dipolar particles with dipole momentum $D$ are assumed
	to be polarized perpendicular to the ring, so that the one-dimensional
	effective long-range DDI is repulsive, and its strength is $g_{d}=D^{2}/\left(4a^{3}_{\bot}\right)$.
	In Eq. (\ref{eq:LongRangeInteraction}), $s=2R\sin\left(\vartheta/2\right)/a_{\bot}$
	is the scaled distance between two particles located at angular positions
	$\theta$ and $\theta^{'}$ in the ring. Noting that the 1D effective
	DDI potential also contains a short-range attractive part $-8g_{d}\delta\left(s\right)/3$
	\citep{sinha_cold_2007,deuretzbacher_ground-state_2010,deuretzbacher_erratum_2013}
	having the same form as the CI potential, we absorb it into the CI
	term. That is to say, the overall CI (we will omit the word ``overall''
	in the following content without causing ambiguity) strength is $g_{0}=-2\hbar^{2}a_{s}/\left(ma^{2}_{\bot}R\right)+8g_{d}a_{\bot}/\left(3R\right)$
	with $a_{s}$ being the s-wave scattering length. In this work, the
	attractive CI, i.e., $g_{0}>0$, will be considered to balance the
	repulsive DDI. We plot the total interaction potential $U\left(\vartheta\right)$
	for a fixed DDI strength and varying CI strength in Fig. \ref{fig:DiagramAndInteractionU}.
	To simplify the formulae, we adopt the natural units $m=\hbar=a_{\bot}=1$
	in the following contents. 
	
	\section{Mean-Field Results}\label{sec:MeanFieldResults}
	\begin{figure}
		\begin{centering}
			\includegraphics{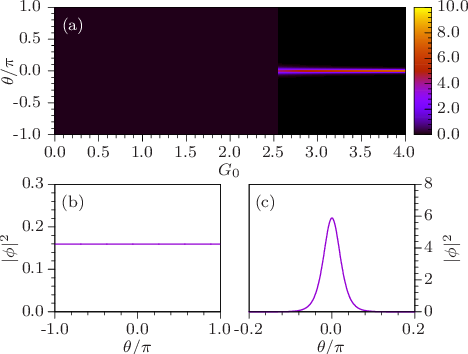}
			\par\end{centering}
		\caption{Mean-field ground state. (a) $\left|\phi\left(\theta\right)\right|^{2}$
			as a function of the CI strength $G_{0}$. (b) Superfluid ground state
			$\left|\phi\left(\theta\right)\right|^{2}$ under the CI strength
			$G_{0}=2.0$. (c) Bright soliton ground state $\left|\phi\left(\theta\right)\right|^{2}$
			under the CI strength $G_{0}=3.0$. The ring trap radius is $R=3.0$,
			and the DDI strength is $G_{d}=3.0$. }\label{fig:MeanFieldGS}
	\end{figure}
	When the system contains a good many particles, it usually can be
	well described within the mean-field theory. In mean-field theory,
	the quantum fluctuation is neglected, and the field operator is replaced
	by its mean value, $\hat{\psi}\rightarrow\phi=\langle\hat{\psi}\rangle/\sqrt{N}$.
	Here, we also scale the field operator by the total particle number
	$N$, so that the macroscopic wave function $\phi$ is normalized
	to one, $\int\left|\phi\right|^{2}d\theta=1$. Under such a replacement,
	the Heisenberg equation for field operator $\hat{\psi}$ becomes the
	time-dependent GPE
	\begin{align}
		i\frac{\partial\phi\left(\theta,t\right)}{\partial t} & =-\frac{1}{2R^{2}}\frac{\partial^{2}\phi\left(\theta,t\right)}{\partial\theta^{2}}-G_{0}\left|\phi\left(\theta,t\right)\right|^{2}\phi\left(\theta,t\right) \nonumber \\ & +G_{d}\phi\left(\theta,t\right)\int\phi^{*}\left(\theta',t\right)U_{d}\left(\vartheta\right)\phi\left(\theta',t\right)d\theta',\label{eq:GPE}
	\end{align}
	where $G_{0}=g_{0}\left(N-1\right)$ and $G_{d}=g_{d}\left(N-1\right)$
	are the particle number scaled CI and DDI strength. Inserting $\phi\left(\theta,t\right)=\phi\left(\theta\right)\exp\left(-i\mu t\right)$,
	with $\mu$ being the chemical potential, into Eq. (\ref{eq:GPE}),
	one obtains a time-independent GPE, and its solution with the minimal
	energy is the ground state of the system. 
	
	The mean-field theory predicts that the ground state of a one-dimensional dipolar
	gas experiences a superfluid to bright soliton phase transition \citep{cuevas_solitons_2009},
	and we show it in Fig. \ref{fig:MeanFieldGS}. In this figure, we
	fix the repulsive DDI strength to $G_{d}=3.0$, and examine the phase
	transition induced by varying the attractive CI strength $G_{0}$,
	which can be experimentally realized by the Feshbach resonance technique
	\citep{chin_feshbach_2010}. When $G_{0}$ is small, the system is
	dominated by the repulsive DDI, which drives the particles to uniformly distribute
	along the ring. Thus, we observe a flat particle density
	profile in this regime [see the left part of panel (a), and panel (b)]. Increasing $G_{0}$ to exceed a critical value
	(in the present figure, it is $G_{\mathrm{0c,mf}}=2.53$), the attractive CI overcomes
	the repulsive DDI, and the net interaction becomes attractive. This
	net attraction, together with the dispersion effect, will lead to
	the formation of a bright soliton [see the right part of panel (a), and panel (c)].
	
	\section{Quantum Results}\label{sec:QuantumResults}
	
	\subsection{Two-Body}\label{subsec:TwoBody}
	We first consider a system of only two dipolar bosons ($N=2$). This
	problem can be reduced to a one-body problem, which can be easily
	solved within the first quantization picture, and provides us with
	a clear physical insight. 
	
	In the first quantization picture, the Hamiltonian
	of two interacting dipolar bosons within a ring trap reads
	\begin{equation}
		H=-\frac{1}{2R^{2}}\left(\frac{\partial^{2}}{\partial\theta^{2}_{1}}+\frac{\partial^{2}}{\partial\theta^{2}_{2}}\right)+U\left(\theta_{1}-\theta_{2}\right),\label{eq:TwoBodyHamiltonian}
	\end{equation}
	where $\theta_{1}$ and $\theta_{2}$ are the angular positions of
	the two particles, and the interaction potential $U\left(\theta_{1}-\theta_{2}\right)$
	has already been given by Eq. (\ref{eq:Interaction}). Introducing
	the center-of-mass coordinate $\Theta=\left(\theta_{1}+\theta_{2}\right)/2$
	and relative-motion coordinate $\vartheta=\theta_{1}-\theta_{2}$,
	this Hamiltonian can be written in a separate-variable form 
	\begin{equation}
		H=H_{c}\left(\Theta\right)+H_{r}\left(\vartheta\right),\label{eq:TwoBodyH_SeparateFrom}
	\end{equation}
	with
	\begin{equation}
		H_{c}=-\frac{1}{4R^{2}}\frac{\partial^{2}}{\partial\Theta^{2}},\quad H_{r}=-\frac{1}{R^{2}}\frac{\partial^{2}}{\partial\vartheta^{2}}+U\left(\vartheta\right).\label{eq:TwoBodyHcAndHr}
	\end{equation}
	The center-of-mass Hamiltonian $H_{c}$ is trivial with ground state
	wave function $\Phi_{c,g}=1/\sqrt{2\pi}$ and ground state energy
	$E_{c,g}=0$. The ground state of the relative motion can be found
	in a standard quantum mechanics procedure. We expand the relative-motion
	wavefunction with a complete set of basis functions, that is
	\begin{equation}
		\Phi_{r}\left(\vartheta\right)=\frac{c_{0}}{\sqrt{2\pi}}+\frac{1}{\sqrt{\pi}}\sum^{K}_{k=1}c_{k}\cos\left(k\vartheta\right).\label{eq:TwoBodyPhirExpansion}
	\end{equation}
	Here, $K$ is a cutoff parameter, and we choose the cosine base functions
	because they are even functions, which automatically fulfill the swap
	symmetry of bosons, $\Phi_{r}\left(\theta_{1}-\theta_{2}\right)=\Phi_{r}\left(\theta_{2}-\theta_{1}\right)$.
	Inserting Eq. (\ref{eq:TwoBodyPhirExpansion}) into the eigenvalue
	equation $H_{r}\Phi_{r}=E_{r}\Phi_{r}$, we obtain a set of linear
	equations of expansion coefficients $\left\{ c_{k}\right\} $, and
	its solution with the minimal eigenenergy is the ground state, which
	we label as $\Phi_{r,g}$. 
	
	\begin{figure}
		\begin{centering}
			\includegraphics{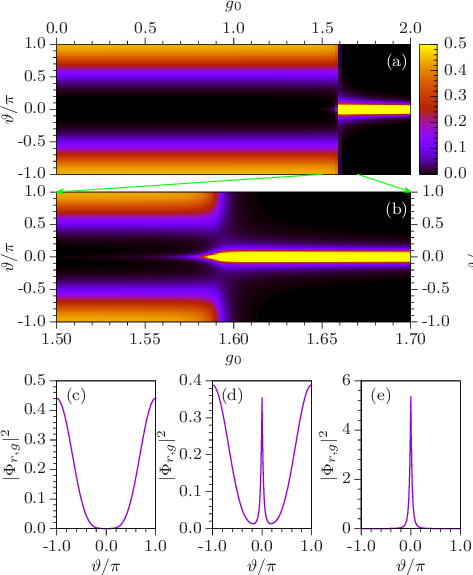}
			\par\end{centering}
		\caption{Two-body relative motion ground state. (a) $\left|\Phi_{r,g}\left(\vartheta\right)\right|^{2}$
			as a function of CI strength $g_{0}$. (b) Enlargement of panel (a)
			in the range of $g_{0}\in\left[1.5,1.7\right]$. (c-e) $\left|\Phi_{r}\left(\vartheta\right)\right|^{2}$
			under CI strength $g_{0}=0$ (c), 1.582 (d), and 1.65 (e), respectively.
			The colorbar is truncated at the upper boundary. In all these panels,
			the ring trap radius is $R=3.0$, and the DDI strength is fixed to
			$g_{d}=3.0$. }\label{fig:TwoBodyPhi_r}
	\end{figure}
	
	The results of the relative-motion ground state are shown in Fig.
	\ref{fig:TwoBodyPhi_r}. In this figure, we fix the DDI strength to
	$g_{d}=3.0$, and examine $\left|\Phi_{r,g}\left(\vartheta\right)\right|^{2}$
	as the CI strength $g_{0}$ increases. When $g_{0}$ is small, $\left|\Phi_{r,g}\right|^{2}$
	is peaked at $\vartheta=\pm\pi$. When $g_{0}$ is large, it is sharply
	peaked at $\vartheta=0$ instead. And, when $g_{0}$ takes a moderate
	value, it is peaked at both $\vartheta=0$ and $\pm\pi$. These results
	can be easily understood by having a look at the interaction potential
	$U\left(\vartheta\right)$, plotted in Fig. \ref{fig:DiagramAndInteractionU}.
	When $g_{0}$ is small, $U\left(\vartheta\right)$ displays a potential
	barrier around $\vartheta=0$; while around $\vartheta=\pm\pi$, it
	exhibits a potential valley. Thereby, $\left|\Phi_{r,g}\right|^{2}$
	is peaked at the potential valley. Increasing the attractive CI strength
	$g_{0}$ will dig a pit around $\vartheta=0$. When the depth of this
	pit is comparable to the potential valley, the total potential exhibits
	a double-well structure, with both $\vartheta=0$ and $\vartheta=\pm\pi$
	being potential valleys. Thus, $\left|\Phi_{r,g}\right|^{2}$is doubly
	peaked there. Further increasing $g_{0}$ will lead to an extremely
	deep pit around $\text{\ensuremath{\vartheta}=0}$, whose attractive
	effect will dominate over the shallow valley around $\vartheta=\pm\pi$.
	As a result, $\left|\Phi_{r,g}\right|^{2}$ is sharply peaked at this
	pit. 
	
	We remark that, unlike $\left|\phi\left(\theta\right)\right|^{2}$
	in the mean-field theory, which is the particle density, here the
	square of the relative-motion wavefunction $\left|\Phi_{r,g}\left(\vartheta\right)\right|^{2}$
	physically means the spatial correlation between the two particles,
	i.e., the probability density of finding the two particles separated
	by an angle of $\vartheta$.  Based on the correlation properties,
	we may identify these three types of states as WC, QD and WC-QD mixed
	states, respectively. For the state with correlation function peaked
	at $\vartheta=\pm\pi$, the two particles have a higher probability
	to separate by an angle of $\pi$, or in other words, they tend to
	crystallize along the ring. Noting that the crystalline phase of electrons
	due to the long-range repulsive Coulomb interaction is usually named
	as Wigner crystal (WC) after its discoverer Eugene Wigner \citep{wigner_interaction_1934,wigner_effects_1938},
	here we may similarly recognize this long-range repulsive DDI induced
	crystal-like state as a WC state. For the state with correlation function
	narrowly peaked at $\vartheta=0$, the two particles tend to stay
	close to each other. We recognize such a self-bounded state as a QD.
	Here, we do not recognize it as a quantum bright soliton because it
	exists at a critical CI strength around $G_{\mathrm{0c,2}}=1.58$,
	which is evidently below the mean-field critical value $G_{\mathrm{0c,mf}}=2.53$
	(see Sec. \ref{sec:MeanFieldResults}). This means that quantum fluctuation
	plays an essential role in the formation of this self-bounded state.
	It induces an effective attractive interaction which can compensate
	for the CI, and lower the transition point. In this sense, we recognize
	this quantum fluctuation induced self-bounded state as a QD rather
	than a quantum bright soliton. We note that the LHY correction also
	predicts a quantum fluctuation induced effective attraction, and the
	formation of QD state in one-dimensional low-density dipolar gas \citep{edler_quantum_2017}.
	At last, the state with correlation function doubly peaked at $\vartheta=0$
	and $\pm\pi$ has both the properties of WC and QD, so we recognize
	it as a WC-QD mixed state.
	
	\subsection{Few-Body}\label{subsec:FewBody}
	
	To deal with the systems containing more than two particles ($N>2$),
	it would be more convenient to employ the second quantization picture.
	We expand the field operator $\hat{\psi}\left(\theta,t\right)$ with
	the angular momentum base states, i.e., 
	\begin{equation}
		\hat{\psi}\left(\theta,t\right)=\frac{1}{\sqrt{2\pi}}\sum_{l}\hat{c}_{l}\left(t\right)e^{il\theta},\label{eq:FieldOperatorExpansion}
	\end{equation}
	where $\hat{c}_{l}$ is the annihilation operator of a boson with
	angular momentum $l$. Then, the Hamiltonian can be expressed in terms
	of $\hat{c}_{l}$, that is 
	\begin{align}
		H & =\frac{1}{2R^{2}}\sum_{l}l^{2}\hat{c}^{\dagger}_{l}\hat{c}_{l}-\frac{g_{0}}{4\pi}\sum_{mnpq}\hat{c}^{\dagger}_{m}\hat{c}^{\dagger}_{n}\hat{c}_{p}\hat{c}_{q}\delta_{m+n,p+q} \nonumber \\ & +\frac{g_{d}}{4\pi}\sum_{mnpq}\hat{c}^{\dagger}_{m}\hat{c}^{\dagger}_{n}\hat{c}_{p}\hat{c}_{q}\mathcal{U}_{d;mnpq},\label{eq:HamiltonianQ2}
	\end{align}
	with 
	\begin{equation}
		\mathcal{U}_{d;mnpq}=\delta_{m+n,p+q}\int e^{i\left(n-p\right)\vartheta}U_{d}\left(\vartheta\right)d\vartheta,\label{eq:ULmnpq}
	\end{equation}
	and $\delta_{i,j}$ being the Kronecker delta function which returns
	1 if $i=j$ and 0 otherwise. We numerically diagonalize the Hamiltonian
	(\ref{eq:HamiltonianQ2}) to find the ground state of the system with
	the help of Julia package QuantumOptics.jl \citep{kramer_quantumopticsjl_2018}.
	Letting $n_{l}$ denote the number of particles in the angular momentum
	state $\psi_{l}=\exp\left(il\theta\right)/\sqrt{2\pi}$, the Hilbert
	space is spanned by the Fock states $|\cdots,n_{-1},n_{0},n_{1},\cdots,n_{l},\cdots\rangle$
	under the restriction of total particle number $\sum_{l}n_{l}=N$
	and zero total angular momentum $\sum_{l}ln_{l}=0$, taking advantage
	of the particle number and angular momentum conservation properties
	of the system. Besides these two conservation restrictions, the Hilbert
	space is also truncated by the non-interacting energy $\sum_{l}n_{l}l^{2}/\left(2R^{2}\right)\leq E_{\mathrm{cut}}$,
	where the cutting energy $E_{\mathrm{cut}}$ is gradually increased
	until the numerical results converge. To explore the quantum nature
	of the system, we mainly examine two physical quantities of the ground
	state --- the normalized two-point correlation function
	\begin{equation}
		g_{2}\left(\vartheta\right)=\frac{\left\langle \hat{\psi}^{\dagger}(\theta)\hat{\psi}^{\dagger}(\theta')\hat{\psi}(\theta')\hat{\psi}(\theta)\right\rangle }{\left\langle \hat{\psi}^{\dagger}(\theta)\hat{\psi}(\theta)\right\rangle \left\langle \hat{\psi}^{\dagger}(\theta')\hat{\psi}(\theta')\right\rangle },\label{eq:g2}
	\end{equation}
	and the condensate fraction $F$, i.e., the largest eigenvalue
	of the reduced single-body density matrix. 
	
\begin{figure}
		\begin{centering}
			\includegraphics{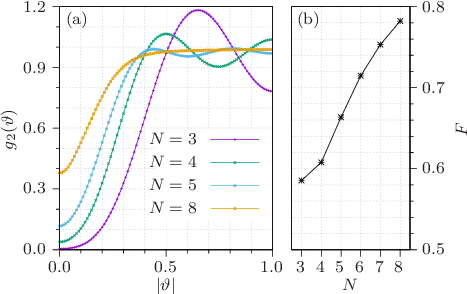}
			\par\end{centering}
		\caption{Two-point correlation $g_{2}$ and condensate fraction $F$ for the
			ground state of different particle numbers. (a) $g_{2}$ as a function
			of $\left|\vartheta\right|$ for different particle numbers. (b) $F$
			as a function of particle number $N$. The ring trap radius is $R=3.0$,
			the DDI strength is $G_{d}=3.0$ and the CI is absent, that is $G_{0}=0$.
		}\label{fig:VaringN}
	\end{figure}

	In Fig. \ref{fig:VaringN}, we show the quantum correlation and condensate
	fraction of the ground state for different particle numbers. In this
	figure, the CI strength is set to zero ($G_{0}=0$); only the repulsive
	DDI plays a role. Due to the repulsion, the probability of finding
	two particles at the same position is suppressed; hence, the two-point
	correlation curve $g\left(\vartheta\right)$ displays a valley around
	$\vartheta=0$, see panel (a). In this panel, for small particle numbers,
	we also observe obvious WC correlation peaks in the region that departs
	from $\vartheta=0$. For example, for particle number $N=4$, the
	green curve is peaked at $\left|\vartheta\right|=\pi/2$ and $\pi$.
	This implies that if we observe one particle at angular position $\theta$,
	then we have a higher probability of finding a second particle at
	positions $\theta'=\theta-\pi/2,\theta+\pi/2$ and $\theta\pm\pi$
	(note that $\theta'=\theta\pm\pi$ are the same position in the ring),
	i.e., the four particles tend to be equidistantly correlated along
	the ring to form a WC state.
	
	In Fig. \ref{fig:VaringN}(a), we also see that as the particle number
	increases, the height of WC correlation peaks decreases and ultimately
	becomes unobservable, leading to a flat correlation plateau in the
	region that departs from $\vartheta=0$, as shown by the yellow line
	for $N=8$. Meanwhile, the correlation valley around $\vartheta=0$
	also becomes shallower as the particle number increases. At the same
	time, Fig. \ref{fig:VaringN}(b) shows that for a small particle number,
	the condensate fraction is low, which implies that quantum depletion
	results in severe fragmentation. Increasing the particle number gives
	rise to a growing condensate fraction. For a system of up to $N=8$
	particles, the condensate fraction already approaches nearly 80\%. According
	to these trends, one would expect that in the $N\rightarrow\infty$
	limit, the system will be a pure condensate with uniform correlation.
	These observations indicate that we can recognize such a state (which
	has a correlation valley around $\vartheta=0$ and a correlation plateau
	otherwise) as SF, and increasing particle number would drive a smooth
	transition from WC to SF state. 
	
	\begin{figure}
		\begin{centering}
			\includegraphics{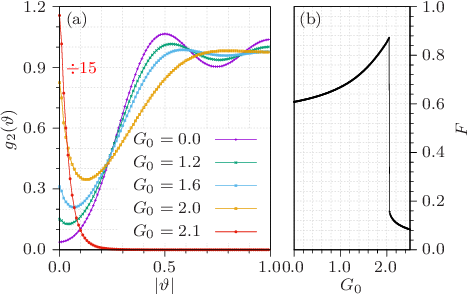}
			\par\end{centering}
		\caption{Two-point correlation $g_{2}$ and condensate fraction $F$ for the
			ground state of different CI strengths. (a) $g_{2}$ as a function
			of $\left|\vartheta\right|$ for different CI strength $G_{0}$. Note
			that the line for $G_{0}=2.1$ is scaled down by a factor of 15. (b)
			$F$ as a function of the CI strength $G_{0}$. The ring trap radius
			is $R=3.0$, the DDI strength is $G_{d}=3.0$, and the particle number
			is $N=4$. }\label{fig:VaringG0_Gd3.0}
	\end{figure}
	
	\begin{figure}
		\begin{centering}
			\includegraphics{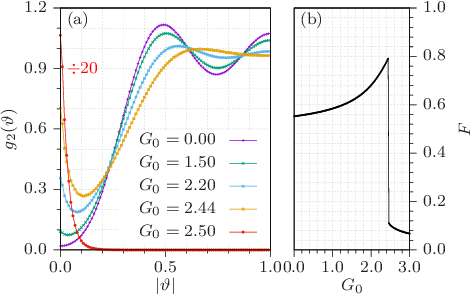}
			\par\end{centering}
		\caption{Same as Fig. \ref{fig:VaringG0_Gd3.0}, except that the DDI strength
			is $G_{d}=4.0$, and in panel (a) the line for $G_{0}=2.5$ is scaled
			down by a factor of 20. }\label{fig:VaringG0_Gd4.0}
	\end{figure}
	
	We also study the effect of CI on the ground state of the system,
	with the results shown in Fig. \ref{fig:VaringG0_Gd3.0}. In this
	figure, the atom number and DDI strength are fixed to $N=4$ and $G_{d}=3.0$,
	while the CI strength varies. When the CI is weak, the ground state
	is a WC state. Increasing the CI strength will reduce the height of
	WC correlation peaks, meanwhile raising a correlation peak at $\vartheta=0$
	which is the attribute of a QD state.
	We recognize this type of state with WC and QD correlation peaks coexisting
	as a WC-QD mixed state. Further increasing the CI strength will flatten
	the WC correlation peaks into a plateau, which is the character
	of an SF state. We recognize this type of state with both SF correlation
	plateau and QD correlation peak as an SF-QD mixed state. During the
	transition from WC to WC-QD mixed to SF-QD mixed state, we observe
	a gradual increase of the condensate fraction, which indicates an
	improvement of the system's coherence. Physically, this is because
	the attractive CI can cancel the effect of repulsive DDI to some degree.
	However, once the CI strength exceeds a critical value (in this figure,
	it is $G_{\mathrm{0c,4}}=2.05$), the condensate fraction quickly
	drops to a very low level, and the correlation becomes sharply peaked
	at $\vartheta=0$. The system begins to enter the complete QD regime
	at this critical point. 
	
	It should be noted that the SF-QD mixed state does not always exist.
	If the DDI becomes stronger, even at the critical value, we can still
	observe the WC correlation peaks. For example, in Fig. \ref{fig:VaringG0_Gd4.0},
	we increase the DDI strength to $G_{d}=4.0$, under which the critical
	CI strength is $G_{\mathrm{0c,4}}=2.44$, and we can still observe
	the WC correlation peaks as shown by the yellow line in panel (a).
	In this case, during the increasing of CI strength, the system skips
	the SF-QD mixed stage, and experiences a transition from WC to WC-QD
	mixed and finally to QD state. 
	
	We further examine the critical CI strength $G_{\mathrm{0c,N}}$ at
	which the transition to QD happens. The critical values are listed
	in Table \ref{tab:CriticalG0} for systems of different particle number
	$N$ and DDI strength $G_{d}$. In this table, the last column lists
	the mean-field critical value at which the SF to bright soliton transition
	happens. From this table, we see that as the particle number $N$
	increases, the quantum critical value gradually approaches the mean-field
	results. This is the result one would expect, because as the particle
	number increases, the effect of quantum fluctuation becomes weak.
	In the $N\rightarrow\infty$ limit, the effect of quantum fluctuation
	is negligible, and mean-field theory can describe the system precisely.
	So, our few-body results are extensible, and could give insights into
	systems with more particles.
	
	\begin{table}
		\begin{centering}
			\begin{tabular*}{8cm}{@{\extracolsep{\fill}}cccccc}
				\hline 
				\noalign{\vskip0.15cm}
				\multirow{2}{*}{$G_{d}$} & \multicolumn{4}{c}{$N$} & \tabularnewline[0.15cm]
				\cline{2-5}
				\noalign{\vskip0.15cm}
				& 2 & 3 & 4 & 5 & Mean-field\tabularnewline[0.15cm]
				\hline 
				\hline 
				\noalign{\vskip0.15cm}
				3.0 & 1.58 & 1.89 & 2.05 & 2.16 & 2.53\tabularnewline[0.15cm]
				\noalign{\vskip0.15cm}
				4.0 & 1.90 & 2.25 & 2.44 & 2.60 & 3.00\tabularnewline[0.15cm]
				\noalign{\vskip0.15cm}
				5.0 & 2.17 & 2.58 & 2.80 & 2.96 & 3.42\tabularnewline[0.15cm]
				\hline 
			\end{tabular*}
			\par\end{centering}
		\caption{Critical CI strength $G_{\mathrm{0c,N}}$ for the transition
			to QD state. $G_{\mathrm{0c,N}}$ for systems of different atom
			number $N=2\sim5$ and different DDI strength $G_{d}=3.0$, 4.0, 5.0
			are listed. The last column lists the mean-field critical value of
			the SF to bright soliton transition. The radius of the ring trap is
			$R=3.0$. }\label{tab:CriticalG0}
	\end{table}
	
	\begin{table}
		\begin{centering}
			\begin{tabular*}{8cm}{@{\extracolsep{\fill}}lcc}
				\hline 
				\noalign{\vskip0.15cm}
				\multirow{2}{*}{States} & \multicolumn{2}{c}{Region}\tabularnewline[0.15cm]
				\cline{2-3}
				\noalign{\vskip0.15cm}
				& Around $\vartheta=0$ & Depart from $\vartheta=0$\tabularnewline[0.15cm]
				\hline 
				\hline 
				\noalign{\vskip0.15cm}
				WC & valley & peaks\tabularnewline[0.15cm]
				\noalign{\vskip0.15cm}
				SF & valley & flat plateau\tabularnewline[0.15cm]
				\noalign{\vskip0.15cm}
				QD & peak & zero\tabularnewline[0.15cm]
				\noalign{\vskip0.15cm}
				WC-QD mixed & peak & peaks\tabularnewline[0.15cm]
				\noalign{\vskip0.15cm}
				SF-QD mixed & peak & flat plateau\tabularnewline[0.15cm]
				\hline 
			\end{tabular*}
			\par\end{centering}
		\caption{Classification of the quantum states by the characteristics of the
			two-point correlation function $g_{2}\left(\vartheta\right)$. For
			details, see the contents in Sec. \ref{sec:Conclusions}. }\label{tab:Classification}
	\end{table}
	
	\section{Conclusions}\label{sec:Conclusions}
	
	In conclusion, we study the quantum states of a few dipolar bosons trapped
	in a ring potential using the exact diagonalization method. Based
	on the characteristics of the two-point correlation function $g_{2}\left(\vartheta\right)$,
	we classify the ground state of the system into five distinct classes:
	WC, SF, QD, WC-QD mixed, and SF-QD mixed states, see Table \ref{tab:Classification}.
	For the WC state with $N$ particles, the correlation exhibits a valley
	around $\vartheta=0$ and $N-1$ peaks nearly equidistantly distributed
	along the ring. The SF state also exhibits a correlation valley around
	$\vartheta=0$; however, in the region that departs from $\vartheta=0$,
	its correlation features a flat plateau with no obvious peaks. In
	contrast, the QD state is characterized by a sharp correlation peak
	at $\vartheta=0$ that quickly drops to zero as $\left|\vartheta\right|$
	increases. The mixed states exhibit a combination of these features:
	the WC-QD mixed state possesses both the QD correlation peak and the
	WC correlation peaks, whereas the SF-QD mixed state combines the QD
	correlation peak and the SF correlation plateau. 
	
	We further examine the transitions between these quantum states. At
	small particle numbers and weak CI, the system exhibits a WC state
	due to the long-range repulsive DDI. Increasing the particle number
	drives a transition from the WC to the SF state. Because increasing
	the particle number weakens the quantum depletion, the condensate
	fraction increases during this transition. Alternatively, increasing
	the CI strength will lead to a sequential transition from the WC state
	to the WC-QD mixed state, to the SF-QD mixed state, and finally to the QD
	state. If the DDI is strong, the stage of SF-QD mixed state may be skipped. In the
	initial WC to mixed state transition stage, the condensate fraction
	increases gradually. However, in the mixed state to the QD state transition
	stage, the condensate fraction sharply drops.
	
	Finally, we compare the exact diagonalization quantum results with
	the mean-field results, showing that as the particle number increases,
	the quantum results gradually approach their mean-field counterparts.
	This behavior suggests that our few-body calculations are robust and
	may help to understand the quantum features of the macroscopic degenerate
	dipolar gas.
	\begin{acknowledgments}
		This work is supported by Guangdong Basic and Applied Basic Research Foundation (Grant No. 2024A1515012526), and National Natural Science Foundation of China (Grant No. 11904063).
	\end{acknowledgments}
	
	\bibliography{refs}
	
\end{document}